# Non-equilibrium Information Envelopes and the Capacity-Delay-Error-Tradeoff of Source Coding

Ralf Lübben, Markus Fidler
Institute of Communications Technology
Leibniz Universität Hannover

*Abstract*—This paper develops an envelope-based approach to establish a link between information and queueing theory. Unlike classical, equilibrium information theory, information envelopes focus on the dynamics of sources and coders, using functions of time that bound the number of bits generated. In the limit the information envelopes converge to the average behavior and recover the entropy of a source, respectively, the average codeword length of a coder. In contrast, on short time scales and for sources with memory it is shown that large deviations from known equilibrium results occur with non-negligible probability. These can cause significant network delays. Compared to well-known traffic models from queueing theory, information envelopes consider the functioning of information sources and coders, avoiding a priori assumptions, such as exponential traffic, or empirical, trace-based traffic models. Using results from the stochastic network calculus, the envelopes yield a characterization of the operating points of source coders by the triplet of capacity, delay, and error. In the limit, assuming an optimal coder the required capacity approaches the entropy with arbitrarily small probability of error if infinitely large delays are permitted. We derive a corresponding characterization of channels and prove that the model has the desirable property of additivity, that allows analyzing coders and channels separately.

## I. Introduction

Originating from the seminal works by Shannon in 1948, the tremendous progress in information and coding theory has enabled numerous ground-breaking applications that range from digital communications to data storage and processing. The fundamental results of information theory are asymptotic limits for the transmission of information by a source over a channel. Information theory defines the notion of entropy and channel capacity as the expected information of a source and the maximum expected transinformation of a channel. Coding theory devises practical source and channel codes for data compression and reliable transmission that seek to approach the limits established by the entropy and the channel capacity, respectively [11].

In networking, information theory has not become widely accepted, yet. A major challenge for establishing a network information theory is due to the properties of network data traffic that is highly variable and delay-sensitive [14]. In contrast, information theory mostly neglects the dynamics of information and capacity and focuses on averages, respectively, asymptotic limits. Typically, these limits can be achieved with arbitrarily small probability of error assuming,

This work has been funded by the German Research Foundation (DFG).

however, arbitrarily long codewords and as a consequence arbitrarily large coding delays [3]. In networking, however, delay is a key performance parameter that can be traded for capacity or loss using results from queueing theory. Moreover, considering the variability of sources is essential in packet data networks as it potentiates significant resource savings due to statistical multiplexing [14].

The analytical cornerstone of networking is queueing theory that dates back to the works on the dimensioning of circuit-switched networks by Erlang in 1909 and 1917. In 1962 Kleinrock advanced the theory and proved the resource efficiency of packet-switching that is achieved by bursty sources due to resource sharing. For packet-switched networks queueing theory can provide exact solutions for backlogs and delays that occur due to the variability of packet inter-arrival and service times. Typically, the inter-arrival and service times obey a certain distribution by assumption, e.g., exponential. Recent approaches like the theory of effective bandwidths [8], [24], deterministic network calculus [8], [12], [25], and the stochastic network calculus [8], [9], [13], [15], [22], [26] compute performance bounds for a wider range of stochastic processes. Despite the need, e.g., for joint coding and scheduling problems or for cross-layer optimization, a tight link between these models and information theory has not been established, so far [3], [14], [22].

To bridge the gap towards queuing theory, a non-equilibrium information theory that can model the variability and delay-sensitivity of real sources is required [3], [14]. While [14] envisions "effective bandwidth versus distortion functions," [3] proposes the idea of "throughput-delay-reliability-triplets" to characterize mobile ad-hoc networks. As potentially promising candidate theories [3], [14], [22] mention effective bandwidths, large deviations, or the stochastic network calculus, however, without providing any details and conclude that unifying information and queueing theory remains as one of the most important challenges.

In this paper we formulate a non-equilibrium theory of information sources and source coders combining methods from information theory and effective bandwidths, respectively, the stochastic network calculus. We characterize information sources by envelope functions that are statistical bounds on the amount of information generated by the source in a time interval of defined width. While on short time-scales the envelopes can exceed the entropy considerably, they approach the entropy on long time-scales and converge in the limit. We derive such information envelopes for memoryless sources and develop a technique for analysis of Markov sources. We find that the memory of a source significantly increases the envelope compared to its entropy and that it leads to a slower convergence. Using a sample path argument for the envelopes we derive a notion of the achievable capacity-delay-error-tradeoff of a coded source. We recover known asymptotic results if the capacity approaches the average codeword length where the delay tends to infinity for any non-trivial probability of error. We show the capacity-delay-error-tradeoff for different coders, including Huffman, Shannon, and Lempel-Ziv. We find that the coder with the smallest average codeword length does not necessarily achieve the best delay performance. We prove that our model has the favorable property of additivity, permitting the independent analysis of sources and channels. We expect that our model enables further joint information- and queueing-theoretical investigations that have the potential to provide substantial new insights and applications from a holistic analysis of communications networks.

The remainder of this paper is structured as follows. In Sec. II we introduce envelope processes and develop the queueing model that we apply in Sec. III to characterize and analyze information sources and coders. In Sec. IV we show how to apply our model to analyze the transmission of coded sources via a Gilbert-Elliott channel and in Sec. V we discuss related works. We provide brief conclusions in Sec. VI.

## II. Envelopes and Performance Bounds

In this section we introduce the concept of statistical envelopes that are the basis of this work. We use the analytical framework of the stochastic network calculus established in [9], [13] to compute statistical performance bounds of the type $\mathsf{P}[\text{backlog} > y] \le \varepsilon$ or $\mathsf{P}[\text{delay} > y] \le \varepsilon$ from envelopes. For a broader overview see, e.g., [17], [22]. In Sec. II-A we develop our model of sources and channels and prove its additivity. In Sec. II-B we assemble a method for construction of statistical envelopes from results on exponentially bounded burstiness [22], [39] and on envelopes [9], [26].

### A. Legendre Transform Model

We use a discrete time model $t \in \mathbb{N}_0$. Denote $A(t)$ the cumulative arrivals at a system, i.e., the cumulative number of bits generated by a source in the interval $[0,t]$. By definition $A(t)$ is a non-negative and non-decreasing random process. By convention $A(0) = 0$. We use shorthand notation $A(\tau,t) = A(t) - A(\tau)$. Similarly, the cumulative departures from a system are denoted $D(t)$. By definition $A(t), D(t) \in \mathcal{F}$ where $\mathcal{F} = \{f : f(t) \ge f(\tau) \ge 0 \ \forall t \ge \tau \ge 0, f(0) = 0\}$.

The service guarantee of a system, e.g., a communications link, a channel, or an entire network, is expressed by a statistical service curve that provides a lower bound for the departures that may be violated with a defined probability. A system has service curve $S(t) \in \mathcal{F}$ with deficit profile $\varepsilon_S(\sigma)$ with $\sigma \ge 0$ if for all $t \ge 0$ it holds that

$$\mathsf{P}[D(t) < A \otimes S(t) - \sigma] \le \varepsilon_S(\sigma) \qquad (1)$$

where $\otimes$ is the min-plus convolution defined for $t \ge 0$ as

$$f \otimes g(t) := \inf_{\tau \in [0,t]} \{f(\tau) + g(t-\tau)\}.$$

Similarly, statistical envelopes provide upper bounds for the arrivals. The arrivals have envelope $E(t) \in \mathcal{F}$ with overflow profile $\varepsilon_E(\sigma)$ with $\sigma \ge 0$ if for all $t \ge 0$ it holds that

$$\mathsf{P}[A(t) > A \otimes E(t) + \sigma] \le \varepsilon_E(\sigma). \qquad (2)$$

Using the definition of service curves and arrival envelopes, statistical backlog and delay bounds can be computed from the maximal vertical and horizontal deviation of $E(t)$ and $S(t)$, respectively.

In this work we use the concave and convex Legendre transforms of $E(t)$ and $S(t)$ defined for $c \ge 0$ as[1]

$$\overline{\mathcal{L}}_E(c) := \sup_{t \ge 0} \{E(t) - ct\},$$

$$\underline{\mathcal{L}}_S(c) := \sup_{t \ge 0} \{ct - S(t)\}$$

to model sources and channels, respectively. Legendre transforms uniquely determine concave arrival envelopes and convex service curves and enjoy a number of useful properties in the network calculus [18]. The following Lem. 1 shows that backlog and delay bounds can be computed from $\overline{\mathcal{L}}_E(c)$ and $\underline{\mathcal{L}}_S(c)$ by a simple addition. The property of additivity is particularly useful as it allows composing results obtained for sources $E(t)$ and systems $S(t)$ independently. Lem. 1 extends an earlier deterministic result for backlogs from [18].

*Lemma 1 (Additivity of Legendre Transforms):* Given a system with service curve $S(t)$ and deficit profile $\varepsilon_S(\sigma)$ and arrivals with envelope $E(t)$ and overflow profile $\varepsilon_E(\sigma)$. For any $c \ge 0$ and $\sigma_E, \sigma_S \ge 0$ it holds for the backlog $B$ that

$$\mathsf{P}[B > \overline{\mathcal{L}}_E(c) + \underline{\mathcal{L}}_S(c) + \sigma_E + \sigma_S] \le \varepsilon_E(\sigma_E) + \varepsilon_S(\sigma_S)$$

and assuming fcfs order it holds for the delay $W$ that

$$\mathsf{P}[W > (\overline{\mathcal{L}}_E(c) + \underline{\mathcal{L}}_S(c) + \sigma_E + \sigma_S)/c] \le \varepsilon_E(\sigma_E) + \varepsilon_S(\sigma_S).$$

Letting $\sigma = \sigma_E + \sigma_S$ we refer to $\varepsilon(\sigma) = \varepsilon_E(\sigma_E) + \varepsilon_S(\sigma_S)$ as the probability of error that can be minimized for $\sigma_E, \sigma_S \ge 0$ as $\varepsilon(\sigma) = \inf_{\sigma_E + \sigma_S = \sigma} \{\varepsilon_E(\sigma_E) + \varepsilon_S(\sigma_S)\} = \varepsilon_E \otimes \varepsilon_S(\sigma)$.

For the special case of a constant rate server with capacity $c$ we have $S(t) = ct$ with $\varepsilon(\sigma) = 0$ for $\sigma \ge 0$ such that $\underline{\mathcal{L}}_S(c) = 0$. It follows from Lem. 1 that $\overline{\mathcal{L}}_E(c) + \sigma_E$ is a backlog bound with probability of error $\varepsilon_E(\sigma_E)$, i.e., $\overline{\mathcal{L}}_E(c)$ has the intuitive interpretation of a backlog bound for arrivals with envelope $E(t)$ at a constant rate server with capacity $c$. Similarly, $\underline{\mathcal{L}}_S(c)$ is a backlog bound for constant rate arrivals with rate $c$ at the system $S(t)$.

*Proof:* Given arrivals $A(t)$ and departures $D(t)$. The backlog of the system is $B(t) = A(t) - D(t)$. By substitution of (1) for $D(t)$ and (2) for $A(t)$ it follows for any $t \ge 0$ that $\mathsf{P}[B > b] \le \varepsilon_E \otimes \varepsilon_S(\sigma)$ where $b = \sup_{t \ge 0} \{E(t) - S(t)\} + \sigma$ [13]. We rewrite $b = \sup_{t \ge 0} \{E(t) - ct + ct - S(t)\} + \sigma$ where $c \ge 0$. It follows that

$$b \le \sup_{t \ge 0} \{E(t) - ct\} + \sup_{t \ge 0} \{ct - S(t)\} + \sigma$$

which completes the proof of the backlog bound.

---
[1]The Legendre transform is also referred to as Fenchel conjugate [32]. Strictly speaking the concave conjugate is defined as $\inf_{t \ge 0} \{ct - E(t)\} = -\sup_{t \ge 0} \{E(t) - ct\}$. We slightly adapt the definition for ease of exposition.

The delay of the system is defined as the horizontal deviation $W(t) = \inf\{\tau \geq 0 : A(t) \leq D(t+\tau)\}$. As above, it follows for any $t \geq 0$ that $\mathsf{P}[W > d] \leq \varepsilon_E \otimes \varepsilon_S(\sigma)$ where $d = \inf\{\tau \geq 0 : \sup_{t \geq 0}\{E(t) - S(t+\tau) + \sigma\} \leq 0\}$. We rewrite

$$d = \inf\{\tau \geq 0 : E(t) - c(t+\vartheta) + c(t+\vartheta) - S(t+\tau) + \sigma \leq 0, \forall t \geq 0\}$$

where $c \geq 0$. We choose $\vartheta = \sup_{t \geq 0}\{E(t) - ct\}/c$ such that $E(t) - c(t+\vartheta) \leq 0$ for all $t \geq 0$ and estimate

$$d \leq \inf\{\tau \geq 0 : c(t+\vartheta) - S(t+\tau) + \sigma \leq 0, \forall t \geq 0\}.$$

After some reordering

$$d \leq \inf\{\tau \geq 0 : c(t+\tau) - S(t+\tau) + \sigma \leq c(\tau - \vartheta), \forall t \geq 0\}$$

we arrive at

$$d \leq \inf\{\tau \geq 0 : (ct - S(t) + \sigma)/c + \vartheta \leq \tau, \forall t \geq 0\}.$$

It follows that $\tau = \vartheta + \sup_{t \geq 0}\{ct - S(t)\}/c + \sigma/c$ and

$$d \leq \sup_{t \geq 0}\{E(t) - ct\}/c + \sup_{t \geq 0}\{ct - S(t)\}/c + \sigma/c$$

completes the proof of the delay bound. ∎

### B. Construction of Envelopes

We construct statistical envelopes as defined in (2) from the moment generating function (MGF) of the arrivals. We assume stationary arrivals, i.e., $\mathsf{P}[A(\tau, \tau+t) > y] = \mathsf{P}[A(t) > y]$ for any $y$ and all $\tau, t \geq 0$. The MGF of the arrivals is

$$\mathsf{M}_A(\theta, t) = \mathsf{E}\big[e^{\theta A(t)}\big]$$

where $\theta$ is a free parameter. Closely related is the concept of effective bandwidths defined for $\theta > 0$ as [8], [24]

$$\alpha(\theta, t) = \frac{1}{\theta t} \ln \mathsf{M}_A(\theta, t). \quad (3)$$

The effective bandwidth increases in $\theta > 0$ from the mean rate of the arrivals in an interval of length $t$ to their peak rate, providing an estimate of their capacity requirements. Given an aggregate of independent arrivals $A(t) = A_1(t) + A_2(t)$ the effective bandwidth $\alpha(\theta, t) = \alpha_1(\theta, t) + \alpha_2(\theta, t)$ is additive, since for the sum of independent random processes it holds that $\mathsf{M}_A(\theta, t) = \mathsf{M}_{A_1}(\theta, t)\mathsf{M}_{A_2}(\theta, t)$.

From Chernoff's theorem $\mathsf{P}[Y \geq y] \leq e^{-\theta y}\mathsf{M}_Y(\theta)$ for $\theta \geq 0$ an upper bound on the arrivals follows as

$$\mathsf{P}[A(t) > F(t) + \varsigma] \leq e^{-\theta(F(t)+\varsigma)}\mathsf{M}_A(\theta, t) = \kappa e^{-\theta\varsigma} \quad (4)$$

where we chose to equate the right hand side with $\kappa e^{-\theta\varsigma}$ with parameters $\kappa \in (0, 1]$ and $\varsigma \geq 0$. We solve for $F(t)$ and obtain

$$F(t) = t\alpha(\theta, t) - \ln\kappa/\theta. \quad (5)$$

By construction $F(t)$ is an envelope for $A(t)$ that is violated at most with probability $\kappa e^{-\theta\varsigma}$ for any $t \geq 0$. It does, however, not satisfy the definition from (2) that requires a sample path argument for all $t \geq 0$. We rewrite (2) as

$$\mathsf{P}[A(t) > A \otimes E(t) + \sigma] = \mathsf{P}[\exists \tau : A(\tau, t) > E(t-\tau) + \sigma] \quad (6)$$

and obtain from the union bound that

$$\mathsf{P}[A(t) > A \otimes E(t) + \sigma] \leq \sum_{\tau=0}^{t-1} \mathsf{P}[A(\tau, t) > E(t-\tau) + \sigma]$$

where we used that the addend at $\tau = t$ is zero since $E(0) + \sigma \geq 0$ and by definition $A(t, t) = 0$.

We select $E(t) = F(t) + \delta t$ where $F(t)$ is given in (5) and $\delta > 0$ is a free parameter. By substitution of $\varsigma = \sigma + \delta t$ we obtain from (4) that $\mathsf{P}[A(t) \geq E(t) + \sigma] \leq \kappa e^{-\theta(\sigma+\delta t)}$ and for $A(t)$ stationary

$$\mathsf{P}[A(t) > A \otimes E(t) + \sigma] \leq \kappa e^{-\theta\sigma} \sum_{\tau=0}^{t-1} e^{-\theta\delta(t-\tau)}.$$

For any $t \geq 0$ we estimate $\sum_{\tau=0}^{t-1} e^{-\theta\delta(t-\tau)} \leq \sum_{\tau=1}^{\infty} e^{-\theta\delta\tau}$. Since $e^{-\theta\delta\tau}$ is decreasing in $\tau$ we can bound each summand by $e^{-\theta\delta\tau} \leq \int_{\tau-1}^{\tau} e^{-\theta\delta\tau}d\tau$ to arrive at

$$\mathsf{P}[A(t) > A \otimes E(t) + \sigma] \leq \kappa e^{-\theta\sigma} \int_0^\infty e^{-\theta\delta\tau}d\tau = \frac{\kappa e^{-\theta\sigma}}{\theta\delta}.$$

Using the definition of envelope (2) we equate $\varepsilon_E(\sigma) = \kappa e^{-\theta\sigma}/(\theta\delta)$. Without loss of generality we choose $\varepsilon_E(0) = 1$ and solve for $\kappa = \theta\delta$ where $\delta \leq 1/\theta$ such that $\kappa \leq 1$. By insertion of $\kappa$ into (5) we derive from $E(t) = F(t) + \delta t$ that $E(t) = (\alpha(\theta, t) + \delta)t - \ln(\theta\delta)/\theta$ has overflow profile $\varepsilon_E(\sigma) = e^{-\theta\sigma}$ and find the Legendre transform

$$\overline{\mathcal{L}}_E(c) = \sup_{t \geq 0}\{(\alpha(\theta, t) + \delta - c)t\} - \frac{\ln(\theta\delta)}{\theta}. \quad (7)$$

For a deterministic constant rate server with capacity $c$ it holds that $\underline{\mathcal{L}}_S(c) = 0$ with deficit profile $\varepsilon_S(\sigma) = 0$ for $\sigma \geq 0$. It follows from Lem. 1 that $\mathsf{P}[B \geq \overline{\mathcal{L}}_E(c) + \sigma] \leq e^{-\theta\sigma}$, i.e., $\overline{\mathcal{L}}_E(c) + \sigma$ is a backlog bound with exponentially decaying probability of error $\varepsilon = e^{-\theta\sigma}$. The parameters $\theta > 0$ and $\delta \in (0, 1/\theta]$ can be optimized to minimize backlog, respectively, delay bounds. Given $\varepsilon$ we can solve $\varepsilon = e^{-\theta\sigma}$ for $\sigma = -\ln\varepsilon/\theta$ and derive the minimal backlog bound

$$b = \inf_{\theta > 0}\left\{\overline{\mathcal{L}}_E(c) - \frac{\ln\varepsilon}{\theta}\right\}.$$

A minimal delay bound follows as

$$d = \inf_{\theta > 0}\left\{\frac{\overline{\mathcal{L}}_E(c)}{c} - \frac{\ln\varepsilon}{\theta c}\right\}. \quad (8)$$

*Remark on Related Envelope Models:* Using the Legendre transform (7) formalizes a backlog bound that can also be derived from the exponentially bounded burstiness model [39] $\mathsf{P}[A(t) > \rho t + \sigma] \leq \kappa e^{-\theta\sigma}$ for $t \geq 0$. By application of the union bound as above a backlog bound for a constant rate server with capacity $c$ is $\mathsf{P}[B \geq \sigma] \leq \kappa e^{-\theta\sigma}/(\theta\delta)$ where $c = \rho + \delta$. Choosing $\kappa = \sup_{\tau \geq 0}\{\mathsf{M}_A(\theta, \tau)e^{-\theta\rho\tau}\}$, that is the optimal solution from Chernoff's theorem, the two backlog bounds can be converted into one another.

We note that a similar result can be obtained by approximation of (6) by the largest term $\mathsf{P}[A(t) > A \otimes E(t) + \sigma] \approx \sup_{\tau \in [0, t]}\{\mathsf{P}[A(\tau, t) > E(t-\tau) + \sigma]\}$ that strictly provides only a lower bound. Letting $E(t) = F(t)$ from (5) where $\overline{\mathcal{L}}_F(c) = \sup_{t \geq 0}\{(\alpha(\theta, t) - c)t\}$ at $\kappa = 1$ yields that $\overline{\mathcal{L}}_F(c) + \sigma$ is a backlog bound that is violated approximately with $e^{-\theta\sigma}$. In comparison, (7) trades the slack rate $\delta$ to achieve a true upper bound.

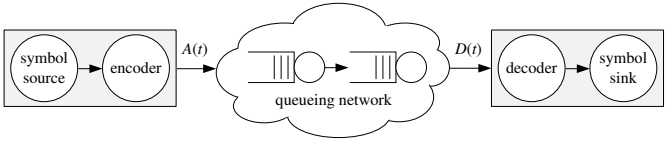

Fig. 1. Unified system model. A source generates symbols according to a defined random process. The symbols are encoded and transmitted as arrivals $A(t)$ by a queueing network. The network departures $D(t)$ are decoded and delivered to the sink.

## III. SOURCE MODELS AND SOURCE CODERS

In this section we investigate the performance of a networked information source. An example of a relevant system is shown in Fig. 1, where the symbols of a source are encoded and transmitted by a network. Our aim is to combine information- and queueing-theoretic aspects to identify achievable operating points within the capacity-delay-error-space of the joint system, i.e., given a network with service curve $S(t)$, e.g., in the most simple case $S(t) = ct$, can the system achieve a delay bound $d$ with probability of error of at most $\varepsilon$?

We specify the detailed system model below. Consider a random variable $X$ that can take any of the values, also called symbols, $x_i$ with probability $p_i$. We also refer to $X$ as the alphabet of the source and denote $|X|$ its cardinality. Information theory defines that if the event $X = x_i$ occurs, it provides information $\mathsf{I}(x_i) = -\operatorname{ld} p_i$ bit where ld denotes the logarithm dualis, i.e., with base 2. The expected information becomes $\mathsf{H}_X := -\sum_i p_i \operatorname{ld} p_i$ that is defined as the entropy of $X$. We label successive symbols generated by a discrete source by $n \in \mathbb{N}$. The stochastic process $X(n)$ has entropy rate $\mathsf{H}_\mathcal{X} = \lim_{n\to\infty} \mathsf{H}(X(1), X(2), \ldots, X(n))/n$, i.e., $\mathsf{H}_\mathcal{X}$ is the entropy per symbol. For stationary processes the entropy rate equals $\mathsf{H}_\mathcal{X} = \lim_{n\to\infty} \mathsf{H}(X(n)|X(n-1), X(n-2), \ldots, X(1))$ [11].

We assign a number of bits $l_i$ to each symbol $x_i$ and define function $l$ to map $x_i$ to $l_i$. Accordingly, $L(n) = l(X(n))$ defines a random process of bit lengths that are generated by the symbol process $X(n)$. As $L(n)$ is an increment process we obtain the cumulative arrival process as $A(n) = \sum_{\nu=1}^{n} L(\nu)$. We let $A(0) = 0$ by definition.

Shannon established the entropy of a source as a fundamental limit for lossless data compression. To this end, a code maps symbols $x_i$ to unique codewords of length $l_i$ where the compression gain is due to assigning short codewords to frequent symbols. If no codeword is a prefix of any other codeword, the code is referred to as a prefix code, where each codeword can be decoded on its own. For an optimal code the expected codeword length $\bar{l} = \sum_i p_i l_i$ is bounded in an interval of one bit width by the entropy as $\mathsf{H}_X \le \bar{l} < \mathsf{H}_X + 1$ [11].

In the next sections we investigate the non-equilibrium behavior of memoryless as well as Markov sources and show examples for finite and infinite alphabets. Secondly, we analyze the performance of well-known coders, such as the Huffman coder, Shannon coder, and Lempel-Ziv coder. Without loss of generality we restrict our investigation to binary codes.

### A. Memoryless Sources

We start our investigation with the basic memoryless source where the symbols $X(n)$ are independent and identically distributed (iid). From the memorylessness it follows that the entropy rate of the process equals the entropy of a single symbol, i.e., $\mathsf{H}_\mathcal{X} = \mathsf{H}_X$. We use function $l$ to assign a number of bits $l_i$ to each symbol $x_i$. By definition $L(n) = l(X(n))$ has categorical distribution with MGF

$$\mathsf{M}_L(\theta) = \sum_i p_i e^{\theta l_i}. \tag{9}$$

For the cumulative arrival process $A(n) = \sum_{\nu=1}^{n} L(\nu)$ it follows that $\mathsf{M}_A(\theta, n) = (\mathsf{M}_L(\theta))^n$ is multinomial. Assuming a source that emits symbols at a constant rate of one symbol per timeslot we substitute $n = t$. We relax this assumption in Sec. III-E. We equate $l_i = -\operatorname{ld} p_i$ such that $\mathsf{M}_A(\theta, t)$ is the MGF of the number of information bits of all symbols generated up to time $t$. From (3) we derive

$$\alpha(\theta) = \frac{1}{\theta} \ln\left(\sum_i p_i^{1-\frac{\theta}{\ln 2}}\right) \tag{10}$$

that does not depend on $t$ due to the memorylessness of the source. An upper envelope on the number of information bits generated by the source up to time $t$ that is violated at most with probability $\kappa$ follows immediately from (5), where $\theta > 0$ is a free parameter that can be optimized.

The envelope provides a benchmark that can be interpreted as a statistical non-equilibrium bound on the number of bits generated by a (hypothetical) optimal coder that maps symbols $x_i$ to codewords of lengths $l_i = -\operatorname{ld} p_i$. The coder is optimal in the sense that it's average codeword length equals the entropy of the source. In practice, this may not be achievable since $-\operatorname{ld} p_i$ typically is non-integer. For comparison, the Shannon code has $l_i = \lceil -\operatorname{ld} p_i \rceil$.

*Geometrically Distributed Symbols:* Assume an infinite alphabet with geometrically distributed symbols $p_i = p(1-p)^i$ for $i \ge 0$. The entropy rate follows by insertion and application of the geometric sum as

$$\mathsf{H}_\mathcal{X} = -\frac{p \operatorname{ld} p + (1-p) \operatorname{ld}(1-p)}{p}.$$

Similarly, $\alpha(\theta)$ follows from (10) for $0 < \theta < \ln 2$ as

$$\alpha(\theta) = \frac{1}{\theta} \ln\left(\frac{p^{1-\frac{\theta}{\ln 2}}}{1 - (1-p)^{1-\frac{\theta}{\ln 2}}}\right). \tag{11}$$

We show respective envelopes from (5) for $p = 0.25, 0.5$, and $0.75$ in Fig. 2. The corresponding entropy rates are $\mathsf{H}_\mathcal{X} \approx 3.25$, $2$, and $1.08$ bit, respectively. The violation probability of the information envelopes is $\kappa = 10^{-6}$. We normalized the envelopes by the corresponding entropy, i.e., we plot $F(t)/\mathsf{H}_\mathcal{X}$. Accordingly, the black line with slope one is the expected normalized information by time $t$. The non-equilibrium information envelopes show a significant deviation from the expected value. The non-linearity of the envelopes arises after minimization of (5) over $\theta > 0$ for any point in time $t \geq 0$. To see the convergence in equilibrium we also depict the increments of the envelopes, that have the interpretation of an information rate, as well as the respective entropy rates $\mathsf{H}_\mathcal{X}$. While the increments of the envelope deviate largely from the entropy on short time scales they converge quickly if longer time intervals are considered.

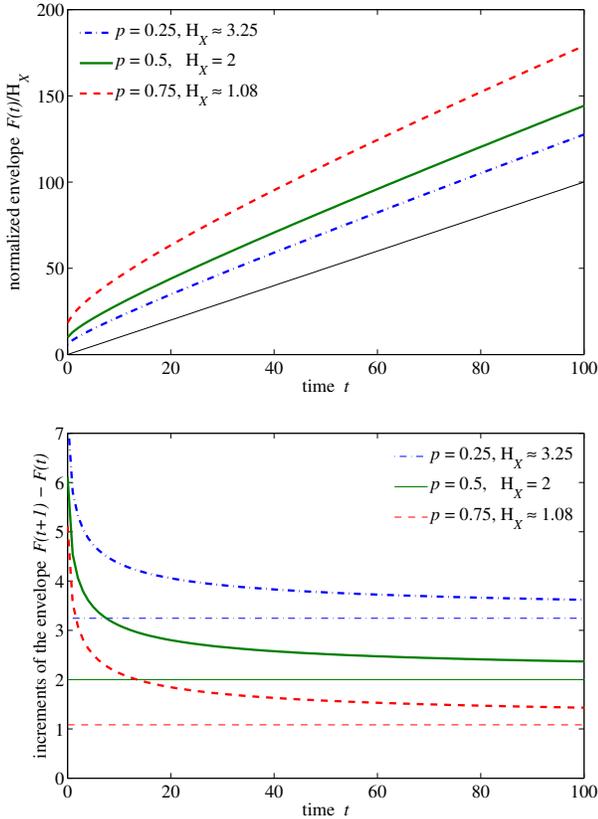

Fig. 2. Information envelopes of a memoryless source with geometrically distributed symbols with parameter $p$. The envelopes show that the actual information rate can be significantly larger than the entropy rate (slope one in the top figure, respectively, horizontal lines in the bottom figure). It converges, however, quickly if longer time intervals are considered.

### B. Huffman Coding

Next, we consider envelopes for the number of bits generated by a Huffman coder and derive performance bounds. To construct a Huffman code execute the following steps repeatedly until all symbols of the source have been processed:

- sort the symbols in decreasing order of probability,
- substitute the two least probable symbols by a new compound symbol, assign the sum of the two probabilities, and add one bit to the respective codewords to distinguish the individual symbols.

The Huffman prefix code achieves the minimal expected codeword length, hence $\mathsf{H}_X \leq \bar{l} < \mathsf{H}_X + 1$. Regarding the individual codeword lengths $l_i$, however, no such simple upper bound exists. In fact, it is shown in [23] that individual codewords of a Huffman code can become as large as approximately $1.44$ times the information of the corresponding symbol, i.e., $l_i < -1.45 \operatorname{ld} p_i$. Compared to the information envelope where $l_i = -\operatorname{ld} p_i$, e.g., Fig. 2, the actual codeword lengths of a Huffman coder may significantly increase the number of bits generated.

We characterize source coders by their capacity-delay-error-tradeoff, i.e., by $(c, d, \varepsilon)$ where $d = \overline{\mathcal{L}}_E(c)/c - \ln \varepsilon/(\theta c)$ for any $\theta > 0$ from (8). Assuming a memoryless source we first obtain $\alpha(\theta)$ for the coder from (9) as

$$\alpha(\theta) = \frac{1}{\theta} \ln \left( \sum_i p_i e^{\theta l_i} \right).$$

Since $\alpha(\theta)$ does not depend on $t$ the condition $c > \alpha(\theta)$ is sufficient to achieve finite $\overline{\mathcal{L}}_E(c)$ from (7). We choose the free parameter $\delta \in (0, 1/\theta]$ as $\delta = c - \alpha(\theta)$. It follows that $\overline{\mathcal{L}}_E(c) = -\ln(\theta\delta)/\theta$ and we obtain from (8) that

$$d = \inf_{\theta > 0} \left\{ \frac{-\ln(\theta(c - \alpha(\theta))\varepsilon)}{\theta c} \right\} \quad (12)$$

is a delay bound with error probability $\varepsilon$.

The $(c, d, \varepsilon)$-tradeoff expresses the capacity that is required to achieve a delay bound subject to a defined probability of error. The delays are due to the randomness that is introduced by variable codeword lengths. Depending on the amount of buffering in the network, the error can be a violation of the delay bound, or a loss of information due to buffer overflow. As an implementation option, the envelopes can be used to discard excess data, that can occur at most with probability $\varepsilon$, proactively by the coder itself, such that the delay bound is not violated. In the limit $\theta \to 0$, i.e., permitting arbitrarily large delays $d \to \infty$ we recover that a capacity of $c \to \bar{l}$ bit per timeslot suffices to transmit the symbols of the source with arbitrarily small probability of error $\varepsilon \to 0$.

*Geometrically Distributed Symbols:* As for Fig. 2 assume an infinite alphabet with geometrically distributed symbols $p_i = p(1-p)^i$ for $i \geq 0$. We let $p = 1/2$ to obtain a dyadic source where $-\operatorname{ld} p_i = i+1$ is integer. The respective Huffman code uses codewords of lengths $l_i = i+1$ such that $\alpha(\theta)$ for the Huffman coder is identical to (11) in this case. Given $c$ and $\varepsilon$ we compute $d$ as described above and optimize $\theta \in (0, \ln(2))$ numerically. The entropy rate of the source is $\mathsf{H}_\mathcal{X} = 2$ and the expected codeword length is $\bar{l} = 2$.

Fig. 3 depicts the $(c, d, \varepsilon)$-tradeoff of the Huffman coded source. For $c > \bar{l}$ finite delay bounds can be computed, whereas the delay grows unbounded for $c \to \bar{l}$. Also, Fig. 3 shows the logarithmic growth of $d$ for decaying $\varepsilon$ that is characteristic of the approach.

### C. Shannon Coding

Shannon coding works as follows. Assume all symbols $x_i$ are ordered in decreasing order of their probabilities, i.e., $p_i \geq p_{i+1}$. Denote $F_i = \sum_{j<i} p_j$ the cumulative probability of all symbols $x_j$ where $j < i$. The first $\lceil -\operatorname{ld} p_i \rceil$ positions after the decimal point of the binary number $F_i$ are the codeword of the symbol $x_i$.

While the Huffman code is optimal with respect to the expected codeword length, certain codewords may exceed the information of the respective symbol significantly. In contrast, the Shannon code achieves codeword lengths $l_i = \lceil -\operatorname{ld} p_i \rceil$ that deviate from the information of any symbol by less than one. While the Shannon code does not generally achieve the minimal expected codeword length it enjoys, however, a property referred to as competitive optimality, i.e., given a randomly selected symbol of a dyadic source, the codeword generated by the Shannon coder is more likely to be smaller than larger if compared to the codeword generated by any other coder [11].

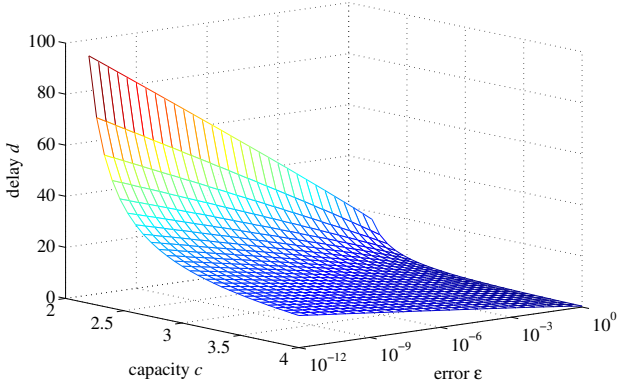

Fig. 3. Capacity-delay-error-tradeoff of a Huffman coded dyadic source with geometric symbol distribution and entropy rate $\mathsf{H}_\mathcal{X} = 2$. The delay grows unbounded if the capacity approaches the expected codeword length $\bar{l} = 2$.

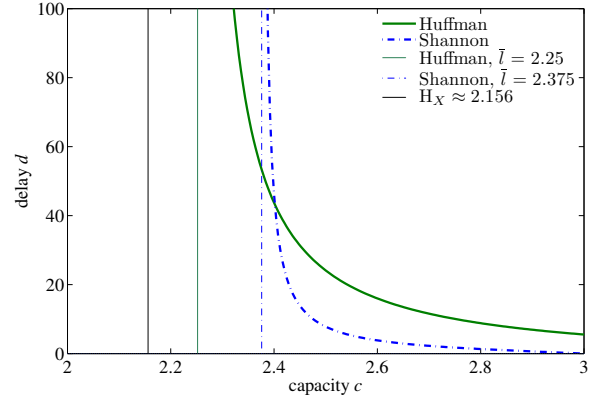

Fig. 4. $(c, d, \varepsilon)$-tradeoff of the Shannon coder compared to the Huffman coder. Unlike the Huffman coder the Shannon coder does not achieve the minimal expected codeword length. Due to the more balanced codeword lengths the Shannon coder effectuates, however, a significantly smaller delay bound if sufficient capacity is available.

To compute the $(c, d, \varepsilon)$-tradeoff we estimate the codeword lengths $l_i = \lceil -\operatorname{ld} p_i \rceil \leq 1 - \operatorname{ld} p_i$. It follows that $\bar{l} \leq \mathsf{H}_\mathcal{X} + 1$ and from (9) $\mathsf{M}_L(\theta) \leq \sum_i p_i e^{\theta(1-\operatorname{ld} p_i)}$ such that for a memoryless source

$$\alpha(\theta) \leq \frac{1}{\theta} \ln\left(\sum_i p_i^{1-\frac{\theta}{\ln 2}}\right) + 1. \qquad (13)$$

As in Sec. III-B we obtain $d$ from (12). Since we compute bounds, we can substitute $\alpha(\theta)$ by the upper bound (13) to obtain a conservative estimate.

Closely related to Shannon coding is Shannon-Fano-Elias coding that motivates arithmetic coding. The respective codewords are of lengths $l_i = \lceil -\operatorname{ld} p_i \rceil + 1$. Using the estimate $l_i \leq 2 - \operatorname{ld} p_i$ the solution follows as above.

*Impact of Codeword Lengths:* Assume a source that has an alphabet of five symbols with probabilities $(3/8, 2/8, 1/8, 1/8, 1/8)$ and $\mathsf{H}_\mathcal{X} \approx 2.156$. The codeword lengths of the Shannon code are $(2, 2, 3, 3, 3)$ such that $\bar{l} = 2.375$. For comparison, the codeword lengths of the Huffman code are $(1, 2, 3, 4, 4)$ with $\bar{l} = 2.25$. We compare the $(c, d, \varepsilon)$-tradeoff of the two coders in Fig. 4 where we optimize the parameter $\theta$ numerically. We choose $\varepsilon = 10^{-6}$ and omit showing further results since $d$ decreases logarithmically with increasing $\varepsilon$ as before, see Fig. 3.

Fig. 4 shows an advantage of the Huffman code compared to the Shannon code if $c \lessapprox 2.4$, that is due to the fact that the Huffman code achieves the minimal expected codeword length, whereas the Shannon code does not. If $c \gtrapprox 2.4$, however, the Shannon code outperforms the Huffman code in terms of the delay due to the smaller variability of the codeword lengths. Since the maximum codeword length of the Shannon code is $\max_i\{l_i\} = 3$ we have $d = 0$ for $c \geq 3$. For the Huffman code we have $\max_i\{l_i\} = 4$ such that $d = 0$ for $c \geq 4$.

### D. Lempel-Ziv Coding

The term Lempel-Ziv coding refers to dictionary-based codes that encode a symbol or a sequence of $s$ symbols by a reference to a previous occurrence. Compared to the codes above, the advantage of Lempel-Ziv coding is that it adapts to the source without a priori knowledge of the symbol distribution. Moreover, it is asymptotically optimal [38]. Here, we consider window-based Lempel-Ziv coding where the window contains the past $w$ symbols. If the current symbol is found in the history, it is replaced by a pointer to the latest occurrence. This coder is proven to be optimal if the sequence length $s$ and the history $w$ tend to infinity [11].

We consider practical implementations with finite $w$ and transmit symbols that cannot be found in the history uncompressed. Assuming a memoryless source $X$ with symbols $x_i$ that occur each with probability $p_i$, the recurrence time $k$ of symbol $x_i$ is geometrically distributed

$$f_i(k) = p_i(1-p_i)^{k-1}.$$

To encode the pointer we use Elias-delta coding that encodes positive integers $j \geq 1$, see [33]. We use the codeword for $j = 1$ that has a length of one bit as a prefix to mark uncompressed symbols and $j \geq 2$ to denote the $k = (j-1)$th most recent symbol in the window. The length of the codewords is [33]

$$l(k) = \lfloor \operatorname{ld}(k+1) \rfloor + 2\lfloor \operatorname{ld}(\operatorname{ld}(k+1)+1) \rfloor + 1$$

and $\alpha(\theta)$ follows from the definition (3) as

$$\alpha(\theta) = \frac{1}{\theta} \ln\left(\sum_i p_i \left(\sum_{k=1}^{w} p_i(1-p_i)^{k-1} e^{\theta l(k)} \right.\right.$$
$$\left.\left. + (1-p_i)^w e^{\theta(\lceil \operatorname{ld}|X|\rceil + l(0))}\right)\right). \qquad (14)$$

The first part of the sum originates from encoding the pointer if the symbol is found in the history. The second part denotes the probability that the symbol does not occur in the history such that the symbol is sent without compression, requiring $\lceil \mathrm{ld}\,|X| \rceil$ bit to encode the symbol where $|X|$ is the cardinality of $X$ plus $l(0) = 1$ bit to mark the symbol as uncompressed. In the sequel we limit the maximum length of encoded pointers to $\lceil \mathrm{ld}\,|X| \rceil$ such that $w = \max\{k : l(k) \leq \lceil \mathrm{ld}\,|X| \rceil\}$.

By insertion of $l(k)$ (14) becomes a sum of polylogarithms such that we cannot provide an analytical solution. For numerical evaluation it is useful to decompose the inner sum to solve $\sum_{k=k_l}^{k_u}(1-p_i)^{k-1}e^{\theta l(k)} = e^{\theta l(k_l)}((1-p_i)^{k_l-1}-(1-p_i)^{k_u})/p_i$ for $k_l = 2^y - 1$, $k_u = 2^{y+1} - 2$ and any $y \geq 1$. As in Sec. III-B we let $\delta = c - \alpha(\theta)$, require $\delta \in (0, 1/\theta]$, and obtain $d$ from (12).

*Impact of the Window Size:* Assume a source has an alphabet of 256 symbols, i.e., an uncompressed symbol uses 8 bit. An overall of 240 of the symbols occur each with probability $1/2048$ and the remaining 16 symbols each with probability $113/2048$. The source generates one symbol per timeslot. The encoder groups $s$ consecutive symbols, causing an additional delay of $s-1$ timeslots, to generate supersymbols with cardinality $|X| = 2^{8s}$. It executes the above algorithm on one of these supersymbols every $s$ timeslots, i.e., the encoder periodically generates a codeword for $s$ symbols every $s$ timeslots. Using the periodicity we can write $\alpha(\theta, t) = \lceil t/s \rceil \ln \mathsf{M}_L/(\theta t)$, where for a single increment $\ln \mathsf{M}_L/\theta$ equals (14). To compute the delay we choose parameter $\delta \in (0, 1/\theta]$ as $\delta = c - \alpha(\theta, 1)/s$. It follows that $\sup_{t \geq 0}\{(\alpha(\theta,t)+\delta-c)t\} = \alpha(\theta,1)(s-1)/s$ and by insertion into (7) and (8) a delay bound is

$$d = \inf_{\theta > 0}\left\{ \frac{\alpha(\theta,1)(s-1)}{cs} - \frac{\ln(\theta(c - \alpha(\theta,1)/s)\varepsilon)}{\theta c} \right\}.$$

Fig. 5 depicts the performance of the Lempel-Ziv encoder for different parameters $s$, see Tab. I. The maximum pointer length is limited to $8s$ bit. The Elias-delta coding of the pointer becomes more efficient with increasing $s$ such that the window size $w$ that can be addressed increases significantly. Accordingly, the probability $p_\mathrm{hit}$ that a random sequence of $s$ symbols is found in $w$ increases. Note that since the algorithm operates on sequences of $s$ symbols, the unit of $w$ are $s$ symbols, too. Finally, the normalized average codeword length $\bar{l}/s$ shows the achievable compression gain. For comparison the entropy rate of the source is $\mathsf{H}_\mathcal{X} \approx 4.98$ and the average codeword length that is achieved by the Huffman coder is $\bar{l} \approx 5.03$ requiring, however, a priori knowledge of the symbol distribution.

Moreover, Fig. 5 shows the $(c, d, \varepsilon)$-tradeoff of the Lempel-Ziv coder for different $s$ compared to the Huffman coder. The capacity requirements of the Lempel-Ziv coder improve with increasing $s$, respectively, increasing window size and approach the entropy eventually. The encoding of sequences of $s$ symbols introduces, however, an additional delay at the encoder. Beyond that, it makes the encoded sequence more bursty, i.e., the encoder emits a codeword for $s$ symbols every $s$ timeslots, which causes further delay.

TABLE I
PARAMETERS OF THE LEMPEL-ZIV CODER.

| $s$ | $w$ | $p_\mathrm{hit}$ | $\bar{l}/s$ [bit] |
|---|---|---|---|
| 1 | $2^4 - 2$ | 0.51 | 7.49 |
| 2 | $2^{10} - 2$ | 0.75 | 7.08 |
| 3 | $2^{16} - 2$ | 0.71 | 6.74 |
| 4 | $2^{24} - 2$ | 0.85 | 6.71 |
| 5 | $2^{31} - 2$ | 0.90 | 6.46 |
| 6 | $2^{38} - 2$ | 0.91 | 6.34 |
| 7 | $2^{46} - 2$ | 0.96 | 6.22 |
| 8 | $2^{54} - 2$ | 0.98 | 6.16 |

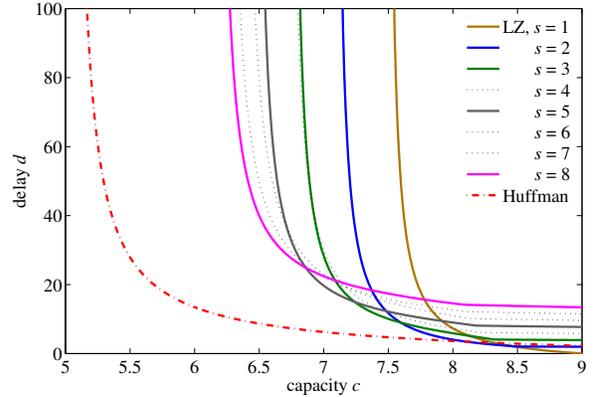

Fig. 5. Lempel-Ziv coding with different window sizes $w$, compared to Huffman coding. The entropy rate of the source is $\mathsf{H}_\mathcal{X} \approx 4.98$. With increasing window size the Lempel-Ziv coder eventually approaches the entropy.

### E. Variable Symbol Rate

So far, we assumed that sources generate symbols at a constant rate. Next, we show how sources with a variable symbol rate can be modeled using conditional MGFs and analyzed by unconditioning. Given a memoryless source and denote $\mathsf{M}_L(\theta)$ the MGF of the increments (9). The conditional MGF of $n$ arrivals becomes $\mathsf{M}_A(\theta, n) = (\mathsf{M}_L(\theta))^n$. Here, the count of arrivals $N(t)$ is a random process with probability mass function $p_N(n, t)$. The MGF $\mathsf{M}_A(\theta, t)$ of the arrival process $A(t)$ follows by unconditioning such that

$$\alpha(\theta, t) = \frac{1}{\theta t}\ln \sum_{n=0}^{\infty}(\mathsf{M}_L(\theta))^n p_N(n, t). \qquad (15)$$

*Poisson Process:* A Poisson process with mean rate $\lambda$ has $p_N(n, t) = e^{-\lambda t}(\lambda t)^n/n!$. By insertion into (15) it follows that

$$\alpha_A(\theta) = \frac{\lambda}{\theta}(\mathsf{M}_L(\theta) - 1)$$

where we used that $\sum_{n=0}^{\infty} a^n/n! = e^a$. Since $\alpha(\theta)$ does not depend on $t$, delay bounds follow immediately from (12).

We show an example for a source that generates eight different symbols with geometrically decreasing probability $p_i = 1/2^i$ for $1 \leq i \leq 7$ and $p_8 = p_7$ such that $\sum_i p_i = 1$. Since the source is dyadic the codeword lengths of the corresponding Huffman code (as well as the Shannon code) are $l_i = -\mathrm{ld}\,p_i$ bit. The entropy rate as well as the average codeword length are $\mathsf{H}_\mathcal{X} = \bar{l} \approx 2$ bit. The MGF of the increments $\mathsf{M}_L(\theta)$ follows from (9). Fig. 6 shows the $(c, d, \varepsilon)$-tradeoff from (12) for the Huffman coded Poisson source. For comparison with this doubly random process, we show results for a Huffman coded constant rate source as well as a hypothetical Poisson source with constant length codewords of length $\mathsf{H}_\mathcal{X}$ bit. The average symbol rate of all sources is $\lambda = 1$. Clearly, the Huffman coded constant rate source achieves zero delay if $c \geq 7$ since the codewords have at most seven bit length, whereas in case of the Poisson arrival process no such limit exists since an arbitrarily large number of symbols may arrive within a single timeslot. Finally, results for an uncoded Poisson source where each symbol is encoded using three bit are shown to depict the compression gain of the Huffman coder.

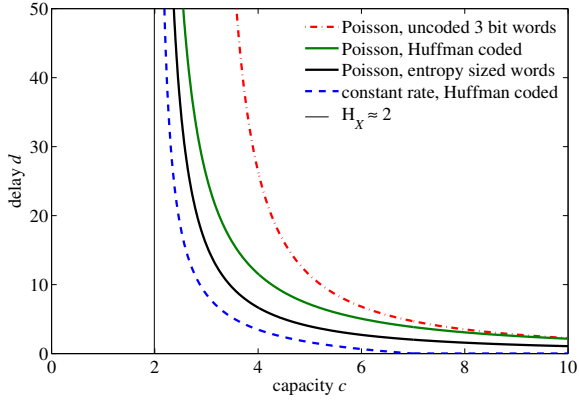

Fig. 6. Huffman coded Poisson source compared to an uncoded Poisson source, a hypothetical Poisson source with constant, entropy-sized codewords, and a Huffman coded constant rate source. The doubly randomness of the Huffmann coded Poisson source causes noticeable delays. Compared to an uncoded Poisson source, the Huffman coder achieves, however, a significant improvement.

### F. Markov Sources

In the following we relax the assumption of memoryless sources and consider discrete, stationary Markov sources, i.e., random processes $X(n)$ with first order dependence where the symbol $x_i$ that occurs in step $n$ depends only on the previous symbol $x_j$ in step $n-1$. The symbol $x_i$ is also referred to as the state of the Markov chain that can take any of the values $i = 1, 2, \ldots, m$. An example of a two state Markov chain is shown in Fig. 7. We denote $p_i$ the stationary state distribution of the chain and $q_{ij}$ the transition probabilities from state $i$ to state $j$. Define $\mathbf{P}$ to be the row vector $(p_1, p_2, \ldots, p_m)$ and $\mathbf{Q}$ to be the state transition matrix. The stationary state distribution is the solution of $\mathbf{P} = \mathbf{PQ}$ under the normalization condition $\mathbf{P1} = 1$ where $\mathbf{1}$ is a column vector of ones.

Due to the first order dependence the entropy rate of a Markov source becomes $\mathsf{H}_\mathcal{X} = \mathsf{H}(X(n)|X(n-1))$ [11] and using the notation above $\mathsf{H}_\mathcal{X} = -\sum_i \sum_j p_i q_{ij} \operatorname{ld} q_{ij}$. Next, we compute information envelopes for Markov sources. The MGF of a discrete Markov chain that produces a constant amount of data $l_i$ if it is in state $i$ is known from [8]. Let $\mathbf{L}$ be the diagonal matrix $\operatorname{diag}(e^{\theta l_1}, e^{\theta l_2}, \ldots, e^{\theta l_n})$. As before we substitute $n = t$ assuming a source that emits symbols

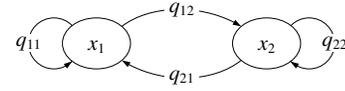

Fig. 7. Example two-state Markov chain.

at a constant rate of one symbol per timeslot. The effective bandwidth of the Markov chain for $t \geq 1$ is known as

$$\alpha(\theta, t) = \frac{1}{\theta t} \ln(\mathbf{P}(\mathbf{L}(\theta)\mathbf{Q})^{t-1}\mathbf{L}(\theta)\mathbf{1}). \quad (16)$$

Regarding (16), we can, however, not substitute $l_i$ by the amount of information generated in state $i$ since the information provided by symbol $x_i$ depends on the previous symbol $x_j$, i.e. each symbol has conditional information $\mathsf{I}(x_i|x_j) = -\operatorname{ld} q_{ji}$ bit. Overall, for a Markov chain with $m$ states we can distinguish $m^2$ distinct pairs of successive symbols.

To solve the problem posed by the conditional information we extend the state space from $m$ to $m^2$ states. We denote the states $i|j$, respectively, $x_i|x_j$ meaning that symbol $x_i$ occurred in the current timeslot after symbol $x_j$ occurred in the previous timeslot. Due to this expansion the information generated by a single symbol in any state of the chain is uniquely determined by the state itself, i.e., the information generated by symbol $x_i$ in state $i|j$ is $\mathsf{I}(x_i|x_j) = -\operatorname{ld} q_{ji}$. The transition probability from state $j|k$ to state $i|j$ is $q_{ji}$ for any $i, j, k$ and zero otherwise. Fig. 8 shows the accordingly extended Markov model for the example from Fig. 7. Given the transition matrix of the extended Markov model we compute the stationary state distribution and let $l_{i|j} = -\operatorname{ld} q_{ji}$ to compute $\alpha(\theta, t)$ from (16). An information envelope follows from (5).

*Two-state Markov Source:* We show an example for a two state Markov source as depicted in Fig. 7. The stationary state distribution follows from the balance equations as $p_1 = q_{21}/(q_{12} + q_{21})$ and $p_2 = q_{12}/(q_{12} + q_{21})$. As a measure of the burstiness of the source we use the average time to change state twice $T = 1/q_{12} + 1/q_{21}$. We choose $p_1 = 5/8$ and $p_2 = 3/8$ and use different burstiness parameters $T \approx 4.3$, $T = 8$, and $T = 16$. The corresponding state transition matrices $\mathbf{Q} = (q_{11}, q_{12}; q_{21}, q_{22})$ are $\mathbf{Q} = (5/8, 3/8; 5/8, 3/8)$, i.e. the source is memoryless, $\mathbf{Q} = (4/5, 1/5; 1/3, 2/3)$, and $\mathbf{Q} = (9/10, 1/10; 1/6, 5/6)$, respectively.

The entropy of a single symbol follows as $\mathsf{H}_X = -\sum_i p_i \operatorname{ld} p_i \approx 0.95$ bit and the entropy rate $\mathsf{H}_\mathcal{X} = \mathsf{H}(X(n)|X(n-1)) = -\sum_i \sum_j p_i q_{ij} \operatorname{ld} q_{ij} \approx 0.95$, $\mathsf{H}_\mathcal{X} \approx 0.80$, and $\mathsf{H}_\mathcal{X} \approx 0.54$ bit, respectively. We use the extended model in Fig. 8 that has the stationary state distribution $p_{1|1} = p_1 q_{11}$, $p_{1|2} = p_2 q_{21}$, $p_{2|1} = p_1 q_{12}$, and $p_{2|2} = p_2 q_{22}$. We compute $\alpha(\theta, t)$ from (16) and information envelopes $F(t)$ from (5) which we minimize for $\theta > 0$.

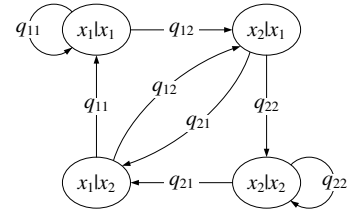

Fig. 8. Extended Markov model for the example from Fig. 7 where the information generated by symbol $x_i$ given the previous symbol was $x_j$ is uniquely determined by the state $x_i|x_j$ itself.

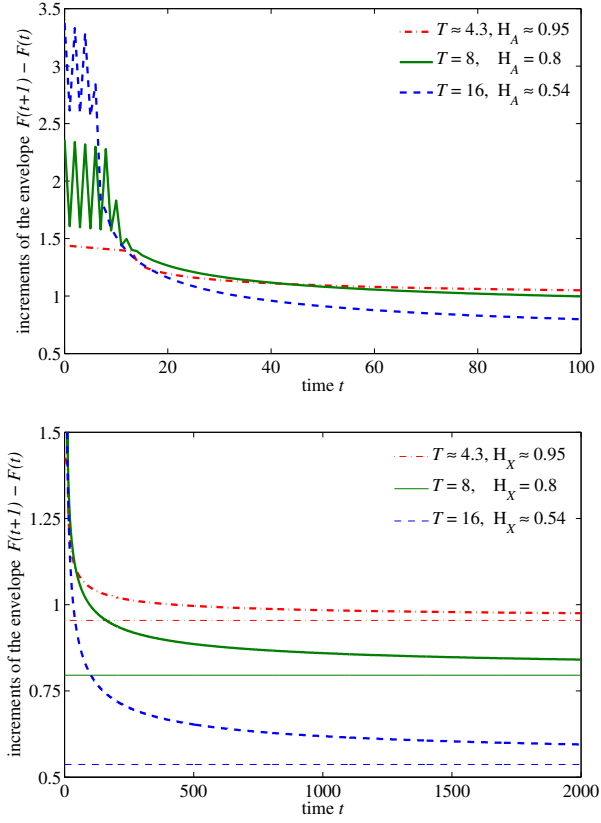

Fig. 9. Increments of the information envelopes of a two-state Markov source with different burstiness parameters $T$, where $T \approx 4.3$ corresponds to a memoryless source. The upper figure zooms into the lower one. While increasing memory $T$ reduces the entropy rate $\mathsf{H}_\mathcal{X}$ it causes a slower convergence of the information envelopes, i.e., the source can deviate significantly from its expected information rate with non-negligible probability.

Fig. 9 shows the increments of envelopes $F(t)$ with violation probability $\kappa = 10^{-6}$. For small $t \lessapprox 10$ the envelopes are determined by the worst-case, i.e., the maximal amount of information that can be emitted by the Markov source. For parameter $T = 8$ the occurrence of symbol $x_2$ after symbol $x_1$ has the largest information $\mathsf{I}(x_2|x_1) = -\mathrm{ld}\, q_{12} \approx 2.32$ bit followed by the occurrence of symbol $x_1$ after symbol $x_2$ with $\mathsf{I}(x_1|x_2) = -\mathrm{ld}\, q_{21} \approx 1.58$ bit. Since direct transitions from state $x_2|x_1$ to state $x_2|x_1$ are not possible, the maximal information is achieved by a sequence of alternating $x_1$ and $x_2$ causing the zigzags in between $\mathsf{I}(x_2|x_1)$ and $\mathsf{I}(x_1|x_2)$ for small $t$. The same argument applies for $T = 16$. In contrast, for $T \approx 4.3$ the source is memoryless such that the information $\mathsf{I}(x_2|x_1) = -\mathrm{ld}\, q_{12}$ equals $\mathsf{I}(x_2|x_2) = -\mathrm{ld}\, q_{22} \approx 1.42$ bit, i.e., the maximum information is achieved by a sequence of all $x_2$ such that zigzags do not occur.

Due to statistical effects for $t \gtrapprox 10$ the worst-case occurs with probability less than $\kappa = 10^{-6}$ such that it does not dominate the envelopes that approach the entropy rate for large $t$. While increasing memory $T$ reduces the entropy rate it causes, however, a significantly slower convergence of the envelope. This is due to unfavorable, high-information sequences of symbols that are not excluded from the envelope by the violation probability $\kappa$.

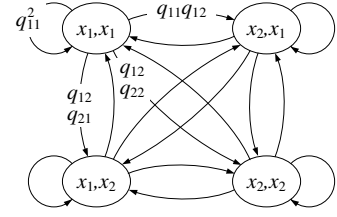

Fig. 10. Extended Markov model for the example from Fig. 7 where states correspond to the occurrence of supersymbols that are sequences of $s$ symbols, here $s = 2$.

### G. Coding Markov Sources

For our investigation of coded Markov sources we assign to each symbol $x_i$ a codeword of length $l_i$ without requiring further assumptions about the coder used. We compute $\alpha(\theta, t)$ of a coded Markov source from (16). To compute a delay bound from (8) we require that $\overline{\mathcal{L}}_E(c)$ from (7) is finite. Since $\alpha(\theta, t)$ increases in $t$ it has to hold that $c > \alpha(\theta, t)$ for all $t \geq 0$. We choose the free parameter $\delta \in (0, 1/\theta]$ as $\delta = c - \sup_{t \geq 0}\{\alpha(\theta, t)\}$. It follows that $\overline{\mathcal{L}}_E(c) = -\ln(\theta\delta)/\theta$ and a delay bound with error probability $\varepsilon$ is

$$d = \inf_{\theta > 0} \left\{ \frac{-\ln(\theta(c - \sup_{t \geq 0}\{\alpha(\theta, t)\})\varepsilon)}{\theta c} \right\}. \quad (17)$$

The compression gain of such a straightforward encoding of a Markov source is, however, limited by the entropy of a single symbol $\mathsf{H}_X$ since the memory of the source is not utilized. To achieve further compression down to the entropy rate $\mathsf{H}_\mathcal{X}$ the coder has to be adapted. One approach is to encode sequences of $s$ symbols instead of single symbols. In this case the average normalized codeword length is limited by $\mathsf{H}(X(1), X(2), \ldots, X(s))/s$ which approaches $\mathsf{H}_\mathcal{X}$ for $s \to \infty$. Given a Markov source with $m$ distinct symbols, respectively, states. If we group $s$ subsequent symbols we can distinguish $m^s$ supersymbols. To model such groups of symbols we extend the state space of the Markov chain to $m^s$ states, accordingly. Fig. 10 shows the extended model for the Markov chain from Fig. 7 for $s = 2$. Here, states $x_i, x_j$, respectively, $i, j$ denote the group of symbol $x_j$ followed by symbol $x_i$. Hence, the state transition probabilities from state $k, y$ to state $i, j$ are $q_{kj} q_{ji}$ for any $i, j, k, y$. We assign a unique codeword to each of the $m^s$ groups of $s$ symbols and use the codeword lengths to determine the diagonal matrix $\mathbf{L}$. For a coder that encodes a group of $s$ symbols every $s$ timeslots $\alpha(\theta, t)$ follows from the extended Markov model for $t \geq 1$ as $\alpha(\theta, t) = \ln(\mathbf{P}(\mathbf{L}(\theta)\mathbf{Q})^{\lceil t/s \rceil - 1}\mathbf{L}(\theta)\mathbf{1})/(\theta t)$. To obtain the delay bound from (8) we choose the free parameter $\delta \in (0, 1/\theta]$ as $\delta = c - \sup_{t \geq 0}\{\alpha(\theta, st)\}$ and compute $\overline{\mathcal{L}}_E(c)$ from (7). Grouping $s$ symbols adds an additional delay of $s-1$ timeslots.

*Two-State Markov Source:* As an example we employ the two-state Markov source as shown in Fig. 7 with transition matrix $\mathbf{Q} = (9/10, 1/10; 1/6, 5/6)$ and encode groups of $s$ symbols using a Huffman coder. Tab. II shows the entropy and the average codeword length normalized by $s$ for $s = 1,\ldots,8$. Clearly, the entropy decreases with increasing $s$ and for $s \to \infty$ we find the entropy rate $H_\mathcal{X} \approx 0.537$. As Tab. II confirms, the Huffman encoding of groups of symbols can approach the entropy rate quite well, however, at the cost of delays. In Fig. 11 we show a delay bound subject to an error probability of $\varepsilon = 10^{-6}$. The delays are due to the variability of the codeword lengths and due to the grouping of $s$ symbols which causes an additional delay of $s-1$ timeslots. Moreover, the grouping makes the encoded sequence more bursty. Depending on $c$ different values of $s$ are optimal, e.g., if $c > 1$ the delay is minimized for $s = 1$ whereas for smaller $c$ larger $s$ are advantageous. Certain parameters $s$, i.e., $s = 3,\ldots,6$ marked by dotted lines, are outperformed for all $c$. This effect is caused by the individual Huffman codes for each $s$ that are more or less efficient.

TABLE II
PARAMETERS OF THE HUFFMAN CODER.

| $s$ | $H(X(1),\ldots,X(s))/s$ [bit] | $\bar{l}/s$ [bit] |
|---|---|---|
| 1 | 0.954 | 1.000 |
| 2 | 0.746 | 0.781 |
| 3 | 0.676 | 0.682 |
| 4 | 0.641 | 0.651 |
| 5 | 0.620 | 0.630 |
| 6 | 0.607 | 0.618 |
| 7 | 0.600 | 0.602 |
| 8 | 0.589 | 0.593 |

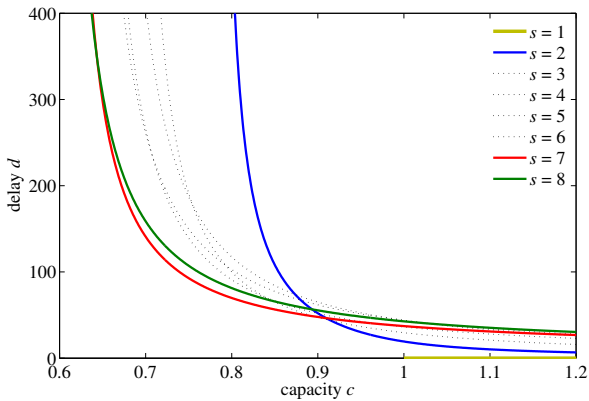

Fig. 11. Huffman coded Markov source. Due to the memory the normalized entropy of groups of $s$ symbols decreases with increasing $s$. The average codeword length of the Huffman code approaches the entropy rate with increasing $s$ resulting, however, in delays due to the variability of the codeword lengths and due to the grouping of symbols.

As an alternative to the grouping of symbols as described above, Markov sources can be encoded efficiently using individual codes for each of the states, i.e., the last symbol determines the code that is used to encode the next symbol. To model an encoder that chooses the code depending on the last symbol we extend the Markov model as described in Sec. III-F, e.g., Fig. 8 for a two-state Markov chain. We denote $l_{i|j}$ the length of the codeword that is used for symbol $x_i$ given the last symbol is $x_j$. Using the extended model $\alpha(\theta, t)$ follows from (16). A delay bound can be computed from (17).

*Example for State Dependent Codes:* Consider a three-state Markov source with transition matrix $\mathbf{Q} = (1/2, 1/4, 1/4; 1/4, 1/2, 1/4; 1/4, 1/4, 1/2)$. We construct an extended nine-state Markov model, as in Sec. III-F, where the code used in state $i|j$ to encode symbol $x_i$ is conditioned on the last symbol $x_j$. Accordingly, if the last symbol was $x_1$, the optimal codeword lengths are $l_{1|1} = 1$ bit, $l_{2|1} = 2$ bit, and $l_{3|1} = 2$ bit, whereas $l_{1|2} = 2$ bit, $l_{2|2} = 1$ bit, and $l_{3|2} = 2$ bit apply if the last symbol was $x_2$, and $l_{1|3} = 2$ bit, $l_{2|3} = 2$ bit, and $l_{3|3} = 1$ bit if the last symbol was $x_3$.

## IV. TRANSMISSION VIA A GILBERT-ELLIOTT CHANNEL

In this section, we show how our results on source coding from Sec. III can be composed with channel models, such as the Gilbert-Elliott channel. Key to this composition is the additivity established by Lem. 1. To this end, we require a service curve model of the channel.

Service curves of wireless channels have been derived, e.g., in [2], [16], [22], [28], [36]. For ease of exposition, we resort to the impairment model from [22]. The model assumes a work-conserving channel, e.g., with peak rate $R$, that is impaired by a stationary random process $I(\tau, t)$. Given $I(\tau, t)$ has envelope $E(t)$ with overflow profile $\varepsilon_E(\sigma)$ (2) the channel has service curve $S(t) = Rt - E(t)$ with deficit profile $\varepsilon_S(\sigma) = \varepsilon_E(\sigma)$ (1) [22].

We assume a two-state Gilbert-Elliott channel that is either in good state, i.e., data are transmitted error-free with rate $R$, or in bad state, i.e., data cannot be decoded and are lost. The transition probabilities between the two states are first order dependent, i.e., the model is a Markov chain. Using the impairment model, the corresponding impairment process is a two-state Markov chain and has rate zero in state 1 (good) and rate $R$ in state 2 (bad) [16], i.e., it consumes no or all available resources, respectively. The effective bandwidth $\alpha(\theta, t)$ of the impairment process is given by (16) and an envelope follows as $E(t) = (\alpha(\theta, t) + \delta)t - \ln(\theta\delta)/\theta$ with $\varepsilon_E(\sigma) = e^{-\theta\sigma}$ where $\theta > 0$ and $\delta \in (0, 1/\theta]$, see Sec. (II-B).

Putting all pieces together, we compute $S(t)$ and obtain the delay bound $(\underline{\mathcal{L}}_S(c) + \sigma_S)/c$ with error probability $\varepsilon_S(\sigma) = e^{-\theta\sigma_S}$ for arrivals with constant rate $c$, where

$$\underline{\mathcal{L}}_S(c) = \sup_{t \geq 0}\{(c + \alpha(\theta, t) + \delta - R)t\} - \frac{\ln(\theta\delta)}{\theta}.$$

As before, we let $\sigma_S = -\ln\varepsilon_S/\theta$ and choose $\delta \in (0, 1/\theta]$ as $\delta = R - c - \sup_{t \geq 0}\{\alpha(\theta, t)\}$ such that $\underline{\mathcal{L}}_S(c) = -\ln(\theta\delta)/\theta$ and a delay bound with error probability $\varepsilon_S$ is

$$d = \inf_{\theta > 0}\left\{\frac{-\ln(\theta(R - c - \sup_{t \geq 0}\{\alpha(\theta, t)\})\varepsilon_S)}{\theta c}\right\}.$$

A delay bound for variable rate arrivals from a source coder follows by a simple addition of the respective Legendre transforms, i.e., from Lem. 1 $(\overline{\mathcal{L}}_E(c) + \underline{\mathcal{L}}_S(c) + \sigma_E + \sigma_S)/c$ is a delay bound for the composed systems with error probability $\varepsilon_E(\sigma_E) + \varepsilon_S(\sigma_S)$.

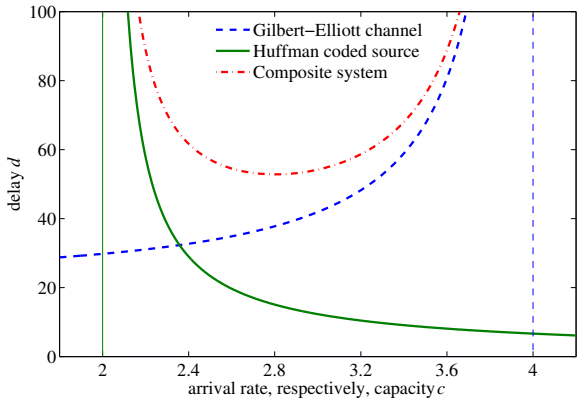

Fig. 12. Transmission of a Huffman coded source via a Gilbert-Elliott channel. The average codeword length of the source is 2 and the average rate of the channel is 4. The individual curves show the delay bound obtained for the Gilbert-Elliott channel given constant rate arrivals with rate $c$, respectively, obtained for the Huffman coded source given a channel with constant service rate $c$. The delay bound for the composite system is obtained from Lem. 1 by taking the minimum of the sum of the two curves, i.e., 53 timeslots.

*Transmission of a Huffman Coded Source:* We consider transmitting the source from Fig. 3 via a Gilbert-Elliott channel with peak rate $R = 6$ and two-state Markov impairment process with generator matrix $\mathbf{Q} = (7/8, 1/8; 1/4, 3/4)$. The state probabilities in equilibrium are $\mathbf{P} = (2/3, 1/3)$ and the average rate of the channel is 4. Fig. 12 shows the individual capacity-delay-error-tradeoffs of the source coder and the channel, each with probability of error $\varepsilon_E = \varepsilon_S = 10^{-6}$. Moreover, we show the sum of the two curves that is a delay bound for the composite system consisting of the Huffman source coder and the Gilbert-Elliott channel for any $c \geq 0$. While $c$ has the interpretation of a constant arrival rate, respectively, constant service rate if we consider $\underline{\mathcal{L}}_S(c)$ and $\overline{\mathcal{L}}_E(c)$ in isolation, it does not have such physical meaning for the composite system, where $\underline{\mathcal{L}}_S(c) + \overline{\mathcal{L}}_E(c)$ can be minimized over $c \geq 0$. The minimal delay bound of the composite system follows as 53 timeslots with probability of error $\varepsilon = 2 \cdot 10^{-6}$.

## V. Related Work

Neglecting the variability and delay sensitivity of real sources, information theory has not become widely accepted in networking so far, see [14] for an excellent survey and a discussion of the gap between respective theories. Recently, [3] proposes non-equilibrium information theory as a new paradigm and highlights the potentialities, difficulties, and possible approaches. The authors envision a characterization of mobile ad-hoc networks by "throughput-delay-reliability-triplets." In this paper we derived a feasible implementation and provided respective models for source coders and channels that complement the vision.

The variability of fading channels is considered already in [29] where a notion of outage capacity is defined. The outage capacity models the probability of errors that occur when the transmission rate is larger than the instantaneous capacity of the channel. A related concept, the delay-limited capacity [21], compensates fluctuations of the fading process using power control to achieve a constant transmission rate. Subsequent works use related concepts to implement power control subject to additional buffering constraints [5], [27]. Recently, the impact of finite blocklength codes on the variability of the channel is investigated, e.g., in [4], [30], [31].

While the definition of outage capacity does not contain any queueing-theoretic aspects, it can be incorporated into a queueing analysis, as shown in [1] using the M|G|1 model. Markovian queues have also been parameterized to model fading channels in [6], [7]. While [6] models a block fading process by a variable rate server that is governed by an embedded Markov chain, [7] views fading outages as an impairment process that is modeled by high priority customers at an M|G|1 priority queue. The concept of an impairment process was also introduced to the stochastic network calculus to analyze outages of wireless channels [22]. Similar to the concept of effective bandwidth [36] develops an effective capacity model to analyze delays due to fading. Multi-access channels are modeled in [34] as a processor sharing queueing system whose capacity is adapted according to the interference created by active stations.

Regarding traffic sources, networking research frequently assumes certain stochastic processes or employs traffic traces. In [20] it is shown how the effective bandwidth of traces, e.g., for MPEG video, can be computed and in [35] empirical envelopes for variable bit rate traffic are derived. The models facilitate performance analysis of networks using respective queueing models. Information theoretic concepts itself are, however, not used. Recent papers [10], [19] provide a framework that includes network elements that process and re-scale data into the analysis. In this work, we model the compression of data by source coders, which complements the approach.

A calculus for so-called information-driven networks is introduced in [37], where the focus is on information instead of data traffic. To this end, the entropy function is employed to convert the data of a flow $A(t)$ to its expected information $\mathsf{H}(A(t))$. By substitution of $\mathsf{H}(A(t))$ for $A(t)$ the framework of the network calculus is used to compute redefined metrics such as the information backlog and the information delay. Compared to [37], in this work we did not define envelopes for the expected information of a source. Instead we derived envelopes for the actual amount of bits generated by memoryless as well as Markov sources and for different implementations of source coders.

## VI. Conclusion

In this paper, we investigated a statistical envelope-based approach towards a non-equilibrium information theory. We applied Legendre transforms to characterize sources and systems by their achievable capacity-delay-error-tradeoff. The additivity of the model facilitates a separability of sources and systems that is comparable to the separation of entropy and channel capacity in information theory. In addition to the average behavior, statistical envelopes and their Legendre transforms consider non-negligible deviations that can cause significant network latencies. If arbitrary delays are permitted, our model recovers the entropy, respectively, average codeword length in the limit. We provided information envelopes for memoryless as well as Markov sources, where we show how the memory increases the variability. We derived the capacity-delay-error-tradeoff of Huffman, Shannon, and Lempel-Ziv coders as well as for Gilbert-Elliott channels. Our models are applicable in the frameworks of the theory of effective bandwidths and the stochastic network calculus enabling joint information- and queueing-theoretic cross-layer research.